\documentclass[nofootinbib]{revtex4}
\usepackage{graphicx}
\usepackage{latexsym}
\def\be{\begin{equation}}
\def\ee{\end{equation}}
\def\bea{\begin{eqnarray}}
\def\eea{\end{eqnarray}}

\begin{document}

\title{Understanding the gravitational and cosmological redshifts as Doppler shifts by gravitational phase factors}

\author{Mingzhe Li}
\email{limz@ustc.edu.cn}
\affiliation{Interdisciplinary Center for Theoretical Study, University of Science and Technology of China, Hefei, Anhui 230026, China}

\begin{abstract}
From the viewpoint of gauge gravitational theories, the path dependent gravitational phase factors define the Lorentz transformations between the local inertial coordinate systems of different positions. With this point we show that the spectral shifts in the curved spacetime, such as the gravitational and cosmological redshifts, can be understood as Doppler shifts. All these shifts are interpreted in a unified way as being originated from the relative motion of the free falling observers instantaneously static with the wave source and the receiver respectively. The gravitational phase factor of quantum systems in the curved spacetime is also discussed.
\end{abstract}

\maketitle

  \vskip 0.4cm

\hfill USTC-ICTS-14-01

\section{Introduction}

As is well known, the received wavelength of a wave propagating in the flat spacetime (without gravity) will get shift if the receiver has a velocity relative to the source.
This is the Doppler effect, a pure kinematical phenomenon. It only depends on the relative velocity of the instantaneous rest inertial frames of the source and the receiver. No dynamical laws are needed.
According to special relativity, the coordinate systems of two inertial frames are related by Poincar\'{e} or inhomogeneous Lorentz transformation, i.e., the homogeneous Lorentz transformation plus a spacetime translation.
The relative motion, due to which the Doppler effect takes place, corresponds to a Lorentz boost.

We also encounter other spectral shifts in the curved spacetime when the gravity is turned on. The most famous examples are the gravitational redshift in the static spacetime and
the cosmological redshift in the expanding universe.  These phenomena are successfully accounted for in general relativity or other gravity theories by the local time dilations due to the difference of the metric at points of the source and the receiver. Except in few works (e.g.,\cite{synge,narlikar,Bunn:2008vj}), they are conventionally thought to be unconnected with Doppler shifts, even though the redshift in the uniform gravitational field was predicted based on the equivalence principle and special relativity in 1907 \cite{Einstein}, before the invention of general relativity. The main objections to interpreting the gravitational and cosmological redshifts as Doppler shifts are based on the following two points: (1) In curved spacetime the concept of relative velocity is ambiguous, especially when the distance between the receiver and the source is large enough to be comparable with the curvature scale of the spacetime; (2) If we naively use the special relativistic Doppler formula with the coordinate velocities to calculate the spectral shifts we will get absurd consequences. For instances, in the case of gravitational redshift if both the source and the receiver are set to be at rest in the spacetime, they have no relative velocity at all and there should be no shift. In the expanding universe if we think the coordinate velocities of the comoving objects are the receding velocities, the Hubble's law implies that distant objects could be receding from us faster than light, in apparent contradiction with relativity \cite{harrison,davis}.

However, we should note that the relative velocities used in the special relativistic Doppler formula are defined between the inertial frames. Both kinds of observers mentioned above, i.e., the rest observers in the static curved spacetime and the comoving observers in the expanding universe, are not inertial. So it is no odd to obtain the contradiction when substituting their coordinate velocities directly into the Doppler formula. The question is whether we can find a right relative velocity to make the Doppler formula valid in calculating the spectral shifts in the curved spacetime. The authors of Refs. \cite{synge,narlikar,Bunn:2008vj} proposed a scheme in which the relative velocity is defined between the local observers by the method of parallel transportation. This scheme can be made more clear as follows. In curved spacetime we don't have any global inertial frames and observers. But Einstein's equivalence principle told us that the small enough regions around any point of the spacetime are indistinguishable from flat Minkowski space, in which the physical laws are the same with those of special relativity and the gravitational effects manifested by the curvature can be neglected. Within these local flat spaces, which are called tangent spaces, the concepts of inertial frames and inertial observers are retained, they are called local inertial frames and free falling observers respectively. Because all measurements are made locally and according to equivalence principle we may get the spectral shift by comparing the wavelengths measured by local free falling observers. For our purpose we should choose the instantaneous static free falling observers in order not to include extra Doppler shifts. In more detail one free falling observer $\mathcal{O}_s$ is instantaneously static with the source at the event of emitting waves and another one $\mathcal{O}_r$ is instantaneously at rest with the receiver at the event of receiving the same wave. The next step is to determine the relative velocity between $\mathcal{O}_r$ and $\mathcal{O}_s$. Since these two observers may be apart from each other by a large distance, it is not possible to compare their velocities directly. But given the affine connections in the spacetime manifold we can parallel transport the four velocity of $\mathcal{O}_s$ to the point of $\mathcal{O}_r$ along the wave's trajectory, then compare it with the four velocity of $\mathcal{O}_r$. In such a way of comparing two velocities at the same point we can get the relative velocity. Substituting it into the special relativistic Doppler formula, the gravitational and cosmological redshifts can be recovered.

In this paper we will revisit this problem by attaching importance on the relation of the local coordinate systems used by free falling observers of different points.
As mentioned earlier in this section, Lorentz boost implies Doppler effect. If we find the transformations between the coordinate systems used by $\mathcal{O}_s$ and $\mathcal{O}_r$ contain a Lorentz boost, we can naturally interpret the corresponding spectral shift as a Doppler shift.

The equivalence principle dictates the existences of local flat spaces and Lorentz transformations within them. This promotes the global Lorentz symmetry in special relativity to a gauge symmetry in curved spacetime and makes it possible to formulate the gravity as a gauge field \cite{utiyama,kibble,hehl}, with the Lorentz group being the gauge group\footnote{Usually in gauge theories of gravity people prefer gauging the Poincar\'{e} group within the framework of Einstein-Cartan theory \cite{kibble,hehl}. In this paper we only discuss the phenomena in the spacetime without torsion, so we consider the case in which only the Lorentz group is gauged.}. The local flat Minkowski space or the tangent space is analogous to the internal space of the conventional gauge theory like Yang-Mills and the local Lorentz transformations are similar to the gauge rotations in the local internal space. In a gauge theory with the gauge group $G$, there are two kinds of $G$-transformations to be distinguished. The first one is the gauge transformation acting on the vectors of internal space at one point, $\psi\rightarrow \psi'=\exp(i\alpha^k T^k)\psi$, here $T^k$ is the generator of $G$ group and $k=1,2,...,m$ with $m$ the dimension of the adjoint representation, $\alpha^k$ is the parameter depending on the spacetime point.
The second one is the non-integrable (path dependent) gauge phase factor associated with a path $\gamma$ from $x_1$ to $x_2$ as showed in \cite{Yang:1974kj}:
\be\label{gaugephase}
U(x_2,x_1)=\mathcal{P} \exp(i\int^{x_2}_{\gamma~x_1} A_{\mu} dx^{\mu})~,
\ee
where $\mathcal{P}$ means path ordering and $A_{\mu}=A_{\mu}^k T^k$ with $A_{\mu}^k$ the gauge potential.
The phase factor for an infinitesimal path from $x$ to $x+\epsilon$ is $U(x+\epsilon, x)=\exp(iA_{\mu}\epsilon^{\mu})=I+iA_{\mu}\epsilon^{\mu}$ and called Wilson link in the lattice gauge theories.
The gauge phase factor (\ref{gaugephase}) plays the role of parallel transporting the basis in the local space
at point $x_1$ to that at point $x_2$, and the gauge potential is the connection. This defined in a non-local way a transformation between these two local spaces. It is important for our purpose that it is an element of the gauge group $G$ since it is generated by the generators $T^k$. Under the gauge transformation, the phase factor changes as
\be
U(x_2,x_1)\rightarrow \exp[i\alpha^k(x_2) T^k]U(x_2,x_1)\exp[-i\alpha^k(x_1) T^k]~.
\ee

In analogy, as we will show in this paper, in the gauge theory of gravity the gravitational phase factors take the role of parallel transporter relating the vectors from different local flat spaces. They are equivalent to transform the vector (for example the wave's momentum measured by a local free falling observer) at one point to that at another point. As same as the conventional gauge theory, the gravitational phase factors are elements of Lorentz group, we may say that the local inertial frames of free falling observers are indeed connected by Lorentz transformations. And we will show in this paper by examples that in the curved spacetime these transformations contain Lorentz boosts and the spectral shifts measured by the instantaneously static free falling observers can be calculated by the special relativistic Doppler formula. The results are identical with those obtained by traditional ways. With this method the spectral shifts in arbitrary curved spacetime can be described in a unified way within the framework of Doppler shift.

We will also extend the discussions to the quantum systems in the curved spacetime.  In this case the gravitational phase factor becomes a quantum Poincar\'{e} transformation in the unitary representation between the wave functions of different points and has the form first obtained by Anandan \cite{anandan1,anandan2} in terms of WKB approximation. It may be used to study the effects of the particle interferometry experiments and to obtain the covariant Schr\"{o}dinger equation of one particle states in the curved spacetime.

This paper is organized as follows. In section II we will briefly review the Lorentz transformations and Doppler effect in special relativity; In section III we will discuss the gravitational phase factor in the gauge theory of gravity; Then apply it to study the kinematical origins of gravitational and cosmological redshifts in section IV;
The gravitational phase factors of quantum systems in the curved spacetime are presented in section V; Finally we conclude in section VI.

\section{Lorentz transformations and special relativistic Doppler effect}

In special relativity a four vector $A^a$ changes as
\be\label{lorentzmatrix}
A^a\rightarrow {\tilde{A}}^{a}=\Lambda^a_{~b}A^b~,
\ee
under the transformation of the inertial coordinate systems, $x\rightarrow \tilde{x}$.
Where $a,~b$ run from $0$ to $3$ and $\Lambda^a_{~b}$ is the Lorentz transformation matrix
which is independent of spacetime points. All the Lorentz transformation matrices satisfy $\Lambda^T \eta \Lambda=\eta$
so that $(\Lambda^0_{~0})^2=1+\Lambda^0_{~i}\Lambda^0_{~i}=1+\Lambda^i_{~0}\Lambda^i_{~0}$ and $|\rm{Det}\Lambda|^2=1$ with $i=1,2,3$,
here we have used the matrix $\eta={\rm diag}(+1,-1,-1,-1)$ to represent the Minkowski metric tensor.
In this paper we only need to focus on the proper orthochronous Lorentz transformations with
\be\label{rapidity}
\Lambda^0_{~0}=\sqrt{1+\Lambda^0_{~i}\Lambda^0_{~i}}~,~|\rm{Det}\Lambda|=+1~,
\ee
which can be obtained from the identity by a continuous change of parameters.
Any other Lorentz transformation can be written as the product of a proper orthochronous one with the space inversion or the time reversal or both.
The matrix elements in Eq. (\ref{rapidity}) can be reexpressed by the rapidity vector $\zeta_i=\zeta \hat{\zeta}_i$ as
\be
\Lambda^0_{~0}=\cosh \zeta~,~~\Lambda^0_{~i}=\sinh \zeta \hat{\zeta}_i~,
\ee
where $\zeta\geq 0$ is the magnitude of $\zeta_i$ which relates the velocity $v_i$ of the frame $x$ relative to $\tilde{x}$ as
follows
\be
\cosh \zeta=\frac{1}{\sqrt{1-v^2}}~,~~\sinh \zeta=\frac{v}{\sqrt{1-v^2}}~,
\ee
and the unit vector $\hat{\zeta}_i$ is identical with the unit vector of the relative velocity $v_i$.

All the proper orthochronous Lorentz transformations can be expressed exponentially as
\be\label{representation}
\Lambda=e^{-\frac{i}{2}\lambda_{ab}J^{ab}}~,
\ee
where the generators $J^{ab}=-J^{ba}$ are $4\times 4$ matrices and have explicit expressions
\be\label{generator}
(J^{ab})^c_{~d}=i(\eta^{ac}\delta^b_{~d}-\eta^{bc}\delta^a_{~d})~.
\ee
Both the generators $J^{ab}$ and parameters $\lambda_{ab}$ are antisymmetric under the exchange of $a$ and $b$. So only six of the generators are independent.
Three of them, $M^i=J^{0i}$, generate the Lorentz boosts and the corresponding parameters are the components of the
rapidity vector $\lambda_{0i}=\zeta_i$. The rest generators, denoted as $J_i={1\over 2}\epsilon_{ijk}J^{jk}$,
generate the spatial rotations
and the parameters $\lambda_{ij}=\epsilon_{ijk}\varphi^k$ are expressed by the rotation angles $\varphi^k$.
With these the Lorentz transformation (\ref{representation}) is rewritten as
\be\label{Lorentz1}
\Lambda=e^{-i(\zeta_iM^i+\varphi^i J_i)}~.
\ee
From Eqs. (\ref{representation}) and (\ref{generator}) it is easy to find that the infinitesimal Lorentz transformations, where $\lambda_{ab}$ are small parameters, have the familiar form
\be\label{inlorentz}
\Lambda^c_{~d}=(1-\frac{i}{2}\lambda_{ab}J^{ab})^c_{~d}=\delta^c_d+\lambda^c_{~d}~.
\ee

All of the proper orthochronous Lorentz transformations form a group, denoted by $SO^{+}(1,3)$. It has representations of varies dimensions. In an $N$-dimensional representation, the element
has the same form of Eq. (\ref{representation})
\be
L(\Lambda)=e^{-\frac{i}{2}\lambda_{ab}\mathcal{J}^{ab}}~,
\ee
where the generators $\mathcal{J}^{ab}$ are $N$-dimensional matrices. All of the representations have the same Lie algebra
\be
[\mathcal{J}^{ab}~,~\mathcal{J}^{cd}]=i (\eta^{bc}\mathcal{J}^{ad}-\eta^{ac}\mathcal{J}^{bd}-\eta^{bd}\mathcal{J}^{ac}+\eta^{ad}\mathcal{J}^{bc})~.
\ee
The Lorentz transformation matrices (\ref{representation}) themselves form a $4$-dimensional representation, this is the fundamental representation of the Lorentz group.
From Eq. (\ref{generator}) one can see that the fundamental representation is not unitary since the boost generators $J^{0i}$ are not hermitian. This is because the Lorentz group is not compact, it has no finite dimensional unitary representations. The generators $J^{ij}$ in Eq. (\ref{generator}) are hermitian. The transformations generated by them form
a compact subgroup, i.e., the rotation group $SO(3)$. Even though the whole Lorentz group has no finite dimensional unitary representations,
it has infinite dimensional unitary representations in which the generators can be chosen to be hermitian.
We will use the notation $U(\Lambda)$ to distinguish the unitary representations from others. Its infinitesimal form is
\be\label{uni}
U(\Lambda)=1-\frac{i}{2}\lambda_{ab}\mathcal{J}^{ab}=1-i(\zeta_i\mathcal{M}^i+\varphi^i \mathcal{J}_i)~.
\ee
The unitary representations take important roles in quantum theory in which the hermitians $\mathcal{J}_i$ are identified with the angular operators of the system.

Only the Lorentz boosts are accompanied by non-vanished rapidities. This will produce the Doppler effect. Taking the electromagnetic waves as the example, if the frame $x$ is instantaneously at rest with the source at the event of emitting light and $\tilde{x}$ is instantaneously at rest with the receiver at the event of receiving the same light, the photon's energy observed by the receiver is
\bea
\tilde{k}^{0}=\Lambda^0_{~0}k^0+\Lambda^0_{~i}k^i~.
\eea
In terms of the null dispersion relation of the photon, the energy ratio should be
\be\label{frequency}
\frac{\tilde{k}^0}{k^0}=(\Lambda^0_{~0}+\Lambda^0_{~i}\hat{k}^i)=\cosh\zeta+\sinh\zeta \hat{\zeta}_i\hat{k}^i~,
\ee
where $\hat{k}^i$ is the unit vector of $k^i$.
The shift of the wavelength is expressed as
\be\label{doppler1}
1+z=\frac{\tilde{\lambda}}{\lambda}=\frac{1}{\cosh\zeta+\sinh\zeta \hat{\zeta}_i\hat{k}^i}~,
\ee
usually $z$ is dubbed redshift, it can be negative when the case actually corresponds to blueshift.
This is the special relativistic Doppler formula.
Let's consider two special cases. In the first case
the source is receding from the receiver, $\hat{\zeta}_i\hat{k}^i=-1$, from Eq. (\ref{doppler1}) one obtains
\be
1+z=e^{\zeta}~,
\ee
or in terms of the relative velocity
\be\label{receding}
1+z=\sqrt{\frac{1+v}{1-v}}~.
\ee
On the contrary, if the source is approaching the receiver, $\hat{\zeta}^i\hat{k}^i=1$ and the blueshift is
\be\label{approaching}
1+z=e^{-\zeta}=\sqrt{\frac{1-v}{1+v}}~.
\ee
In both cases the redshift can be expressed in a unified way
\be\label{redshift}
1+z=e^{-\zeta_i\hat{k}^i}~.
\ee

\section{Gravitational Phase Factor in Lorentz Gauge Theory of Gravity}

In the curved spacetime, inertial frames and coordinate systems only have local meanings.
At any point the local inertial coordinate systems can be constructed in terms of the tetrad (or vielbein), which is a set of orthonormal basis vectors $\hat{e}_a$ and relates to the natural coordinate basis of the tangent space $\partial_{\mu}$ in the following way
\be
\hat{e}_a=e^{\mu}_{~a}\partial_{\mu}~.
\ee
The dual basis in the cotangent space is a set of one-forms denoted as
\be
\hat{\theta}^{a}=e_{\mu}^{~a}dx^{\mu}~.
\ee
Here we meet two sets of indices denoting the tensorial components: the spacetime indices $\mu, \nu, ...$ and the local space indices $a, b, ...$, both of them run from $0$ to $3$. The spacetime indices are raised and lowered by the spacetime metric $g_{\mu\nu}$ and the local
space indices are raised and lowered by the Minkowski metric $\eta_{ab}$.
The components of the tetrad vectors satisfy the relations
\be
g_{\mu\nu}e^{\mu}_{~a}e^{\nu}_{~b}=\eta_{ab}~,~~~\eta_{ab}e_{\mu}^{~a}e_{\nu}^{~b}=g_{\mu\nu}~.
\ee
Any tensor can be expressed in terms of its components in the spacetime coordinate basis or in the local basis. For example a vector $A$ is written as $A=A^{\mu}\partial_{\mu}=\bar{A}^a\hat{e}_a$.
From now on we will use the bar to denote the components read from the local coordinate systems.
The choice of the tetrad basis at each point is arbitrary, we may choose other tetrad bases at the same point. This is the gauge freedom.
All of the bases at a fixed point $p$ relate to each other by the local Lorentz transformations defined on $p$, i.e., the gauge transformations,
\be\label{inverselorentz}
\hat{e}'_{a}(p)=(\Lambda^{-1})^b_{~a}(p)\hat{e}_b(p)~,
\ee
where $(\Lambda^{-1})^b_{~a} (p)=\eta_{ac}\eta^{bd}\Lambda^{c}_{~d}(p)$ is the inverse Lorentz transformation since it acts on the basis. Consequently the components of a vector change as
\be
\bar{A}'^{a}(p)=\Lambda^a_{~b}(p)\bar{A}^b(p)~.
\ee

The local flat spaces at different points are independent. The vectors from different local spaces can not be compared directly. However, when the affine connections are given, the relations among these local spaces can be constructed.
First the relations among the tetrad bases of different points can be seen from the so-called ``tetrad postulate" \cite{carroll}, which states that the covariant derivative of the tetrad field vanishes
\be
D_{\mu}e_{\nu}^{~a}\equiv\partial_{\mu} e_{\nu}^{~a}-\Gamma^{\rho}_{~\mu\nu}e_{\rho}^{~a}+\omega_{\mu~b}^{~a}e_{\nu}^{~b}=0~,
\ee
where $\Gamma^{\nu}_{~\mu\rho}$ is the Levi-Civita connection and $\omega_{\mu~b}^{~a}$ is the spin connection, it has the property $\omega_{\mu ab}=-\omega_{\mu ba}$ to guarantee the orthonormality of the tetrad basis at each point. This means the tetrad basis is parallel transported along any path $x(\eta)$ and $\eta$ is the affine parameter,
\be\label{parallel}
\frac{dx^{\mu}}{d\eta}D_{\mu}e_{\nu}^{~a}=0~.
\ee
This equation means the tetrad $e_{\nu}^{~a}$ at any point can be thought to be come from a parallel transport of the tetrad at another point along arbitrary path.
For our purpose, we need to know the propagation of the wave in the spacetime. The electromagnetic wave propagates along the light curve, a null geodesic on the manifold and is denoted as $x(\eta)$. The photon's momentum $k^{\mu}=dx^{\mu}/d\eta$ is the tangent vector of the geodesic and automatically parallel transported along $x(\eta)$,
\be\label{trans}
\frac{dx^{\mu}}{d\eta}D_{\mu}k^{\nu}=\frac{dx^{\mu}}{d\eta}(\partial_{\mu}k^{\nu}+\Gamma^{\nu}_{~\mu\rho}k^{\rho})=0~.
\ee
Combining this equation with Eq. (\ref{parallel}), we obtain that the local components of the momentum, $\bar{k}^a(\eta)=k^{\nu}(\eta)e_{\nu}^{~a}(\eta)$, measured by the free falling observers who are instantaneously static with the comoving observers along the light curve also satisfy the parallel transport equation,
\be
 \frac{dx^{\mu}}{d\eta}D_{\mu}\bar{k}^a=\frac{dx^{\mu}}{d\eta}D_{\mu}(k^{\nu}e_{\nu}^{~a})=0~.
 \ee
 According to the definition of the covariant derivative, above equation implies
\be
\frac{dx^{\mu}}{d\eta}(\partial_{\mu}\bar{k}^a+\omega_{\mu~b}^{~a}\bar{k}^b)=0~.
\ee
From which we know that the momenta observed by a pair of instantaneous static free falling observers at neighboring points
$x(\eta)$ and $x(\eta+d\eta)=x(\eta)+dx$ are related to each other as follows
\be\label{infi}
\bar{k}^a(\eta+d\eta)=(\delta^a_b-\omega^a_{~b}) \bar{k}^b(\eta)~,
\ee
where $\omega^a_{~b}=\omega_{\mu~b}^{~a} dx^{\mu}$ is the spin connection one form.
The transformation matrix element $\delta^a_b-\omega_{~b}^{a}$ in above equation
is nothing but the infinitesimal Lorentz transformation (\ref{inlorentz}) mentioned in section II with the infinitesimal parameter being just the negative spin connection one form, $\lambda^a_{~b}=-\omega^a_{~b}$. So it can be rewritten as
\be
\bar{k}^a(\eta+d\eta)=[e^{\frac{i}{2}\omega_{cd}J^{cd}}]^a_{~b}\bar{k}^b(\eta)\equiv \Lambda^a_{~b}(\eta+d\eta, \eta)\bar{k}^b(\eta)~.
\ee
The transformation matrix
\be
\Lambda(\eta+d\eta, \eta)=e^{\frac{i}{2}\omega_{cd}J^{cd}}
\ee
is the gravitational phase factor associated with an infinitesimal path, similar to Wilson link in lattice gauge theories. Its eigenvalues are not really phase factors with unity modulus because it is not a unitary matrix.
It takes the role of parallel transporter to relate the vector in the local tangent space at one spacetime point to that at another point. In analogy to the conventional gauge theory, the spin connection is called gauge potential.

Now consider a set of instantaneous rest free falling observers along the photon's trajectory, each one is away from her adjacent partners by a small interval $dx$, then her measured momentum of the photon will
change from that measured by her former partner by an infinitesimal Lorentz transformation as Eq. (\ref{infi}). So the transformation from the momentum measured by $\mathcal{O}_r$ to that by $\mathcal{O}_s$
is an accumulation of infinite number of the infinitesimal transformations,
\be\label{pathLorentz}
\bar{k}^a(\eta_r)=\Lambda^a_{~b}(\eta_r,~\eta_s) \bar{k}^b(\eta_s)~,
\ee
and the transformation matrix is
\be\label{Lorentz}
\Lambda(\eta_r,~\eta_s)=\exp[{\frac{i}{2}\omega_{c_1d_1}(\eta_1)J^{c_1d_1}}]\exp[{\frac{i}{2}\omega_{c_2d_2}(\eta_2)J^{c_2d_2}}]\cdot\cdot\cdot
\exp[{\frac{i}{2}\omega_{c_Nd_N}(\eta_N)J^{c_Nd_N}}]~,
\ee
where $\eta_1=\eta_r,~\eta_N=\eta_s$ and $\eta_1>\eta_2>\cdot\cdot\cdot >\eta_N$ with $N\rightarrow \infty$.
It can be rewritten in a more abstract form
\be\label{Lorentz2}
\Lambda(\eta_r,~\eta_s)=\mathcal{P}\exp[{\frac{i}{2}\int_{s}^{r}\omega_{\mu cd}(\eta)J^{cd}}dx^{\mu}(\eta)]~.
\ee
This gravitational phase factor for a finite path is also an element of Lorentz group, it is
non-locally defined and different from the local Lorentz transformation defined only at one point. Since this is a Lorentz transformation,
any spectral shift observed by free falling observers can be understood to be Doppler shift. This is consistent with the equivalence principle because the free falling observer $\mathcal{O}_r$ is not aware of the existence of gravity or spacetime curvature, she will interpret the observed spectral shifts as being originated from the relative velocity of $\mathcal{O}_s$.
Under a gauge transformation, the phase factor transforms as
\be
\Lambda(\eta_r,~\eta_s)\rightarrow \Lambda(\eta_r)\Lambda(\eta_r,~\eta_s)\Lambda^{-1}(\eta_s)~,
\ee
where $\Lambda(\eta_r)$ is the local Lorentz transformation at the point $x(\eta_r)$ and $\Lambda^{-1}(\eta_s)$ is the inverse Lorentz transformation at $x(\eta_s)$.
In the following section we will use these results to calculate the gravitational and cosmological redshifts with Doppler formula.

\section{Kinematical origins of gravitational and cosmological redshifts}

\subsection{Gravitational Redshift}

In general relativity the gravitational redshift refers to the spectral shift taking place in the stationary spacetime, for example the region outside a spherical and massive object, which is described by the Schwarzschild line element
\be
ds^2=g_{00}(r)dt^2-\frac{1}{g_{00}(r)}dr^2-r^2(d\theta^2+\sin^2\theta d\phi^2)~,
\ee
with $g_{00}=1-2GM/r$ and $M$ is the mass of the object. Both the source and the receiver are static and have fixed coordinate $r,\theta,\phi$.
Because
\be
ds^2=\eta_{ab}\hat{\theta}^a\hat{\theta}^b~,
\ee
the dual basis can be chosen as
\be\label{dualbasis}
\hat{\theta}^0=\sqrt{g_{00}}dt,~\hat{\theta}^1=\frac{dr}{\sqrt{g_{00}}},~\hat{\theta}^2=rd\theta,~\hat{\theta}^3=r\sin\theta d\phi~.
\ee
Such a choice picks up the instantaneous rest free falling observers. This can be seen from the following argument. Eq. (\ref{dualbasis}) told us that $e_{\mu}^{~0}=\sqrt{g_{00}}\delta^0_{\mu},~e_{\mu=0}^{~i}=0$. A static observer has the four velocity $U^0=1/\sqrt{g_{00}},~U^i=0$ in the spacetime coordinate system. In the local basis $\bar{U}^a=e_{\mu}^{~a}U^{\mu}=\delta^a_0$, so $\bar{U}^0=1$ and $\bar{U}^i=0$.
This means the static observer has no spatial velocity in the local inertial frames chosen above.

Using the Maurer-Cartan structure equation
\be
d\hat{\theta}^a+\omega^a_{~b}\wedge \hat{\theta}^b=0~,
\ee
where $d\hat{\theta}^a$ means exterior differentiation and $\wedge$ is the wedge product, it is easy to find out the non-vanished spin connection one-forms as listed below
\bea
& &\omega^0_{~1}=\omega^1_{~0}={1\over 2}g_{00}' dt~, \omega^1_{~2}=-\omega^2_{~1}=-\sqrt{g_{00}} d\theta~,\nonumber\\
& &\omega^1_{~3}=-\omega^3_{~1}=-\sqrt{g_{00}} \sin\theta d\phi~,\omega^2_{~3}=-\omega^3_{~2}=-\cos\theta d\phi~,
\eea
where $g_{00}'\equiv d g_{00}/dr$.
In order to simplify the discussion, we consider the photon transports along the radial direction, so that on the light curve the angles $\theta$ and $\phi$ are constants, only $\omega^0_{~1}$ and $\omega^1_{~0}$ are relevant. In this case each
exponential term on the right hand side of Eq. (\ref{Lorentz}) commutates with each other and we obtain
\be
 \Lambda(\eta_r,~\eta_s)=\exp[{i\int_{s}^{r}\omega_{\mu 01}(\eta)J^{01}}dx^{\mu}(\eta)]
 =\exp[{\frac{i}{2}\int_{s}^{r} g_{00}'(\eta)M^1}dt]~.
\ee
Comparing the above expression with Eq. (\ref{Lorentz1}), we find that the above equation represents the boost along the $\hat{e}_1$
direction with the rapidity
\be
\zeta_1=-{1\over 2}\int_{s}^{r} g_{00}'dt~.
\ee
We require both the source and the detector are outside the Schwarzschild radius, $r_r,~r_s>2GM$.
Using the line element of the light ray $ds^2=0$, one has
\be
dt=\pm\frac{dr}{g_{00}}~,
\ee
where the plus corresponds to the case $r_r>r_s$ in which photons propagate outwards and ${\hat{\bar{k}}}^1=1$ because the positive direction of $\hat{e}_1$ chosen in Eq. (\ref{dualbasis}) is the same as the radial direction with increasing radius. The minus sign corresponds to the case $r_r<r_s$ in which photons propagate inwards and ${\hat{\bar{k}}}^1=-1$.
With these we have in both cases
\be
 \zeta_1{\hat{\bar{k}}}^1= -{1\over 2}\int_{s}^{r} d\ln g_{00}= \ln\sqrt{\frac{g_{00}(\eta_s)}{g_{00}(\eta_r)}}~.
\ee
Then using the Doppler formula (\ref{redshift}), we obtain the redshift
\be
1+z=\exp({-\zeta_1{\hat{\bar{k}}}^1})=\exp(-\ln\sqrt{\frac{g_{00}(\eta_s)}{g_{00}(\eta_r)}})=\sqrt{\frac{g_{00}(\eta_r)}{g_{00}(\eta_s)}}~.
\ee

This result is identical with that got from the traditional way in general relativity. Up to now we have shown that the gravitational spectral shift can be understood to be originated from the relative motion between local free-falling observers who do not feel the existence of gravity. Even though the source is static with the receiver in the spacetime coordinate system, their instantaneously rest free falling partners have a relative velocity. And the free falling observers will interpret the spectral shift as a Doppler shift. One can also use Eqs. (\ref{receding}) and (\ref{approaching}) to derive the relative velocities.
When $r_r>r_s$, $g_{00}(\eta_r)>g_{00}(\eta_s)$ and consequently $z>0$, this corresponds to a redshift, the observer $\mathcal{O}_r$ will see that $\mathcal{O}_s$ is receding from her. On the contrary, if $r_r<r_s$, it is a blueshift, $\mathcal{O}_r$ will see that $\mathcal{O}_s$ is approaching her.
This can be understood intuitively as follows. In the first case, the photon propagates outwards the gravitational center $r=0$. First the free falling observer $\mathcal{O}_s$ is static with the spacetime, then fall towards the gravitational center.
When the photon arrives $r_r$, $\mathcal{O}_s$ has already obtained a velocity towards $r=0$, but at this moment $\mathcal{O}_r$ is at rest with the spacetime so she will see that $\mathcal{O}_s$ is receding from her.
In the second case, the directions of the photon's propagation and $\mathcal{O}_r$'s falling velocity are the same, at the event of receiving $\mathcal{O}_r$ will see that $\mathcal{O}_s$ is approaching her.

\subsection{Cosmological Redshift}

Similarly we can interpret the cosmological redshift as a Doppler shift from the free falling observers' views. In the expanding Friedmann-Robertson-Walker (FRW) universe, the line element is
\be
ds^2=dt^2-a^2(t)[\frac{dr^2}{1-Kr^2}+r^2d\theta^2+r^2\sin^2\theta d\phi^2]~,
\ee
where $a(t)$ is the scale factor and $K$ is the curvature of the universe.
The dual basis for the local inertial frames may be constructed as
\be
\hat{\theta}^0=dt,~\hat{\theta}^1=\frac{adr}{\sqrt{1-Kr^2}},~\hat{\theta}^2=ard\theta,~\hat{\theta}^3=ar\sin\theta d\phi~.
\ee
It is easy to check that the free-falling observers with above bases are instantaneously rest with the comoving observers. In this spacetime we have the following independent non-vanished spin connection one-forms
\be
\omega^0_{~i}=H \hat{\theta}^i~, ~\omega^1_{~2}=-\sqrt{1-Kr^2}d\theta~,~\omega^1_{~3}=-\sqrt{1-Kr^2}\sin\theta d\phi~,~
\omega^2_{~3}=-\cos\theta d\phi~,
\ee
where $H=d\ln a/dt>0$ is the Hubble expansion rate.
Similarly, for simplicity we consider the case in which photons propagate radially and along the light curve only $\omega^0_{~1}=\omega^1_{~0}=\frac{aHdr}{\sqrt{1-Kr^2}}$ are non-zero.
Putting the receiver to the origin of the coordinate, $r_r=0$, so we always have $r_s>r_r$ and ${\hat{\bar{k}}}^1=-1$. With the null line element $ds^2=0$, we have
\be
\frac{dt}{a}=-\frac{dr}{\sqrt{1-Kr^2}}~,~\omega_{01}=\omega^0_{~1}=-Hdt=-d\ln a~.
\ee
In terms of these equations, one can obtain
\be
\Lambda(\eta_r,~\eta_s)=\exp[i\int^r_s\omega_{01}M^1]
 =\exp[-i(\int^r_s d\ln a)M^1]=\exp[-i\ln(\frac{a_r}{a_s}) M^1]~.
\ee
Hence the rapidity is
\be
\zeta_1=\ln(\frac{a_r}{a_s})~,
\ee
and finally we have the redshift
\be
1+z=e^{-\zeta_1{\hat{\bar{k}}}^1}=e^{\zeta_1}=\frac{a_r}{a_s}~.
\ee
This is the famous formula for the cosmological redshift, here it is interpreted by the Doppler shift caused by the relative motion between
the local inertial observers at different positions. Similarly we can use Eq. (\ref{receding}) to deduce the relative velocity of the free falling observer $\mathcal{O}_s$ to $\mathcal{O}_r$. Of course this relative velocity is different from the coordinate velocity of the comoving object and it can never exceed the speed of light.

In fact this procedure can be generalized to arbitrary smoothed curved spacetime with given metric tensor $g_{\mu\nu}$.
From the line element
\be
ds^2=g_{\mu\nu}dx^{\mu}dx^{\nu}=\eta_{ab} \hat{\theta}^a\hat{\theta}^b~,
\ee
we can choose the cotangent basis
\be\label{basis}
 \hat{\theta}^0=\sqrt{g_{00}}dt+\frac{g_{0i}}{\sqrt{g_{00}}}dx^i~,~\hat{\theta}^k=e_{\mu=i}^{~k}dx^i~,
 \ee
 where the tetrad components are determined by
 \be
e _{\mu=0}^{~0}=\sqrt{g_{00}}~,~e _{\mu=0}^{~k}=0~,~e _{\mu=i}^{~0}=\frac{g_{0i}}{\sqrt{g_{00}}}~,~e _{\mu=i}^{~k}e _{\mu=j}^{~k}=\frac{g_{0i}g_{0j}}{g_{00}}-g_{ij}~.
 \ee
This choice picks up the free falling observers instantaneously at rest with the observers who have fixed spatial coordinates $x^k$ and four velocities $U^0=1/\sqrt{g_{00}}$ and $U^k=0$. Using the above tetrad components, it is easy to see that at each point these velocities are expressed in the local basis as $\bar{U}^a=\delta^a_0$. With the chosen basis (\ref{basis}), we can use the Cartan structure equation to compute the spin connection one forms. Then substitute the spin connections to Eq. (\ref{Lorentz2}) and use the property of null geodesic $\eta_{ab}\hat{\theta}^a\hat{\theta}^b=0$ to get the Lorentz transformation between $\mathcal{O}_r$ and $\mathcal{O}_s$. From it we may read out the rapidity vector, then use the Doppler formula to obtain the redshift $1+z$.

\section{Some Thoughts on Gravitational Phase Factors of Quantum Systems in Curved Spacetime}

The key point to understand the spectral shifts in curved spacetime as the Doppler shifts is that the vectors from different local spaces are connected by the non-locally defined Lorentz transformations: the gravitational phase factors. These transformations depend on the paths connecting the end points. As far as the redshift experiment is concerned, the path is fixed, it is the trajectory of the wave (photon's geodesic in the examples of this paper), so the transformation is also fixed.

Up to now, both the gravitational phase factor and the local Lorentz transformation in our considerations act on the four dimensional vector spaces, they are in the fundamental representation of the Lorentz group. Similar discussions may be extended to other dimensional representations. Now we attempt to generalize previous discussions to the quantum systems in the curved spacetime. As we know, for a special relativistic quantum system (described by the quantum field theory), its state vectors or wave functions form a physical Hilbert space. The Lorentz transformations of inertial observers' coordinates will induce unitary transformations on the Hilbert space and all of these unitary transformations furnish an infinite dimensional representation of the Lorentz group, this is the unitary representation (\ref{uni}). In the curved spacetime, the inertial observers and the concept of Lorentz invariance can only have local meanings. We may envisage that the Hilbert space of a relativistic quantum system is also defined locally. This means that at each point of the spacetime we have a Hilbert space, which are isomorphic to but independent of each other. The set of Hilbert spaces at all points form a Hilbert bundle over the spacetime manifold. The state vector field (or the wave function over the spacetime) is the cross-section of the bundle. These concepts are analogous to the tangent spaces and tangent bundle and have been proposed or mentioned before in the literature \cite{Prugovecki:1994ru,Drechsler:1995ay,anandan2}. According to the equivalence principle, at each local Hilbert space, the dynamics obeys the special relativistic quantum laws. Lorentz symmetry is stated as: at a fixed point $p$ if an inertial observer sees a system in a state $\Psi$, another inertial observer at the same point will see that it is in an equivalent but not the same state $U(\Lambda(p))\Psi$, where $U(\Lambda(p))$ is an element of Lorentz group in the unitary representation Eq. (\ref{uni}), which depends on the point $p$.

This picture is different from the traditional quantum field theory in curved spacetime \cite{davies} where the Heisenberg picture is used and the wave function is dependent of the spacetime. It also differs from the functional representation approach \cite{jackiw} in which the wave function is associated to the field configuration on the space-like hypersurface at each moment. Whether this picture can be as successful as or consistent with the other ways remains unclear. But it seems to the author that it is also a natural way to combine the quantum field theory in flat space and the equivalence principle.

In this picture, the state vectors at different points live in different local Hilbert spaces. When the affine connections are given, they are related by the gravitational phase factors which parallel transport the state vector seen by a free falling observer at one point to that seen by a free falling observer at another point. Naively these phase factors are elements of the Lorentz group in the unitary representation. However, there are some differences from the case we met in studying the redshift of the classical waves in previous sections. First in quantum physics the concepts of particles' trajectories such as the geodesics are not clear, so the path along which the state vectors are parallel displaced can be chosen arbitrarily. Different paths lead to different results, which are related by gauge transformations. Second, the origins of the local inertial coordinate systems changed from point to point and the changes of origins are identical to the spacetime translations which also affect the wave functions. In the discussions on the redshifts of classical waves, the four momenta of photons are unaffected by the spacetime translations. So we should take into account the whole Poincar\'{e} group in the quantum case. For nearby points the local inertial coordinate systems are related by an infinitesimal Poincar\'{e} transformation
\be
\xi^a(x+dx)=(\delta^a_b-\omega^{~a}_{\mu~b}dx^{\mu})\xi^b(x)+e_{\mu}^{~a}(x)dx^{\mu}=(\delta^a_b-\omega^{a}_{~b})\xi^b(x)+\hat{\theta}^a(x)~,
\ee
where the displacement $dx^{\mu}$ in the curved spacetime corresponds to $e_{\mu}^{~a}(x)dx^{\mu}=\hat{\theta}^a(x)$ in the local inertial frame, the state vectors seen by the free falling observers are related to each other by
\be\label{di}
\Psi_{x+dx}=
[1+\frac{i}{2}\omega_{ab}(x)\mathcal{J}^{ab}-i\hat{\theta}^a(x)P_a]\Psi_x~,
\ee
where $P_a$ is generator of the translation and this equation describes how the wave function change with the spacetime points.
Using the same procedure we get the state vectors of the quantum system seen by the free falling observers away by a finite distance as follows
\be\label{pathintegral}
\Psi_{\eta_f}=\mathcal{P}\exp[i\int_{i}^{f}({1\over 2}\omega_{\mu ab}\mathcal{J}^{ab}-e_{\mu}^{~a}P_a)dx^{\mu}(\eta)]\Psi_{\eta_i}~,
\ee
the result depends on the path $x(\eta)$ and the unitary operator $\mathcal{P}\exp[i\int_{i}^{f}({1\over 2}\omega_{\mu ab}\mathcal{J}^{ab}-e_{\mu}^{~a}P_a)dx^{\mu}(\eta)]$ is the gravitational phase factor which has been obtained in Refs. \cite{anandan1,anandan2} within WKB approximation. In this case the gravitational phase factor becomes a operator on the Hilbert spaces. As before it is an element of Poincar\'{e} group but non-locally defined.

The equation (\ref{pathintegral}) may be used to calculate the phase shifts measured in the particle interferometry experiments such as the Colella-Overhauser-Werner (COW) experiment \cite{Colella:1975dq} where the neutron's phase shift due to the gravity has been observed.
Consider neutrons transport along two paths $\gamma_1$ and $\gamma_2$ (e.g., the classical trajectories of neutrons) from one point $\eta_i$ to another point $\eta_f$, if the first transportation results in the wave function $\Psi_1$, the wave function $\Psi_2$ resulted from the second transportation should be a local Lorentz transformation of $\Psi_1$ at the point $\eta_f$,
\be
\Psi_2=U[\Lambda(\eta_f)] \Psi_1~.
\ee
The interferometer at $\eta_f$ will detect the interference fringe due to the superposition $\Psi_1+\Psi_2$.
We can also see from Eq. (\ref{pathintegral}) that besides the Lorentz boost and the translation the rotation also contributes to the resulted wave function in general case. The term $\omega_{\mu ab}\mathcal{J}^{ab}$ along the path is the coupling of the rotated frame (e.g., the rotation of the Earth) to the total angular momentum. For neutrons which have spin, the total angular momentum is the sum of the
orbit angular momentum and the spin, $\mathcal{\vec{J}}=\vec{L}+\vec{S}$. Hence in addition to the effect arising from the coupling of the frame's rotation to orbital angular momenta of particles, which is called Sagnac effect and has been observed in experiments \cite{Werner:1979gi,Atwood}, there should be also contribution from the rotation-spin coupling as proposed in Ref. \cite{Mashhoon:1988zz} and studied extensively in the literature, e.g. \cite{Hehl:1990nf,Cai:1991wj}.

The equation (\ref{di}) implies the generalized Schr\"{o}dinger equation for the quantum system in the curved spacetime
\be\label{dii}
i\partial_{\mu}\Psi=(e_{\mu}^{~a}P_a-{1\over 2}\omega_{\mu ab}\mathcal{J}^{ab})\Psi~.
\ee
There is another understanding of this equation from the viewpoint of covariantization of the Sch\"{o}dinger equation in the gauge theory of gravity.
In the flat space quantum field theory, the energy-momentum operator generates the spacetime translation to the system
\cite{Weinberg:1995mt} and it can be represented as 
\be\label{s}
i\partial_a\Psi=P_a\Psi~,
\ee
where $P_0=H$ is the Hamiltonian and $P_i=-P^i$ is the momentum operator. For a free field, $P_aP^a=m^2$ is the Casimir invariant. An example of the form of the Hamiltonian can be taken for the scalar field $H=(1/2)\int d^3k k^0[\hat{a}(\vec{k})\hat{a}^{\dag}(\vec{k})+\hat{a}^{\dag}(\vec{k})\hat{a}(\vec{k})]$, where $k^0=\sqrt{k^2+m^2}$ and $\hat{a},~\hat{a}^{\dag}$ are annihilating and generating operators respectively. 
However, in curved spacetime the equation (\ref{s}) is not covariant under the local Lorentz transformation $\Psi\rightarrow U[\Lambda(x)]\Psi$. In order to get the gauge covariant
Schr\"{o}dinger equation, we must replace the ordinary derivative $\partial_a$ with the covariant one $D_a$. It is not difficult to find that
\be
D_a=e^{\nu}_{~a}(\partial_{\nu}-{i\over 2}\omega_{\nu cd}\mathcal{J}^{cd})~,
\ee
and
\be
iD_a\Psi=P_a\Psi~.
\ee
Multiply the above equation with $e_{\mu}^{~a}$ we will get Eq. (\ref{dii}). We want to stress that the above equation is Schr\"{o}dinger equation about the evolution of wave function. 
This is different from other first order equations in the quantum field theory.  For example the Dirac equation governs the evolution of the fermion fields which are field operators and spacetime dependent in the Heisenberg picture.  
One immediate result of Eq. (\ref{dii}) or its integral form (\ref{pathintegral}) is that the vacuum state annihilated by all the Poincar\'{e} generators is spacetime independent.
This is different from the traditional approach to the quantum field theory in curved spacetime \cite{davies} where the concepts of vacuum state and particles are vague.

\section{Conclusions}

In this paper we revisited the problem of understanding the redshifts in curved spacetime within the frame of Doppler shifts from the viewpoint of gauge theory of gravity.
We note that the local inertial coordinate systems at different spacetime points are related by the gravitational phase factors, which
are non-locally defined Lorentz transformations (or Poincar\'{e} transformations when the changes of the coordinate origins are considered). Since we can transform the wave's momentum measured by the instantaneous static free falling observer at the source to that at the receiver by a Lorentz transformation as appeared in special relativity, we can treat the spectral shifts taking place in the curved spacetime on the same footing with the Doppler effects. Making use of this method we obtained the correct results for the gravitational and cosmological redshifts. This interpretation suggests that the shifts are originated from the relative motion of the free falling observers. We also discussed the gravitational phase factors of quantum systems in curved spacetime and its theoretical and experimental implications. Our discussions are based on suggesting the existence of the Hilbert bundle, which is quite similar to the tangent bundle in the classical case. The latter is a natural consequence of the equivalence principle. The concept of Hilbert bundle can also be considered as a result of combination of the equivalence principle and quantum mechanics.

In both classical and quantum cases, we need to distinguish two kinds of Lorentz transformations carefully. The first one is the local Lorentz transformation at a fixed point, which is similar to the gauge rotation taking place in the internal space in the conventional gauge theories. The second is the gravitational phase factor relating two vectors from different local spaces. It is a non-locally defined parallel transporter. Its analogue in the conventional gauge theories is the gauge phase factor or Wilson line. It mapped an element of the gauge group (it is the Lorentz group in this paper) to a path in the spacetime manifold. These two kinds of transformations take different roles in physics. The first one expresses a symmetry and the second one dictates the interaction, for which the gauge fields are needed. Mathematically the phase factor has the form of symmetry. Just for this reason we are able to interpret the spectral shifts in curved spacetime which is essentially produced by the gravitational coupling, in terms of their kinematical analogous, i.e., the Doppler shifts.

\section{Acknowledgement}

The author is grateful to J. X. Lu for stimulating discussions and reading the manuscript.
This work is supported in part by Program for New Century Excellent Talents in University and by National Science Foundation of China under Grants No. 11075074.

{}

\end{document}